\documentclass[prb,onecolumn, preprint, superscriptaddress]{revtex4-1}

\usepackage{graphicx}  % needed for figures
\usepackage{bm}        % for math
\usepackage{amssymb}   % for math

\begin{document}

\title{Ellipticity dependence of high-harmonic generation in solids: \\
 unraveling the interplay between intraband and interband dynamics}

 \author{Nicolas Tancogne-Dejean}
  \email{nicolas.tancogne-dejean@mpsd.mpg.de}
  \affiliation{Max Planck Institute for the Structure and Dynamics of Matter, 
  Luruper Chaussee 149, 22761 Hamburg, Germany}
 \affiliation{European Theoretical Spectroscopy Facility (ETSF)}%, Luruper Chaussee 149, 22761 Hamburg, Germany}

  \author{Oliver D. M\"ucke}
  \affiliation{Center for Free-Electron Laser Science CFEL, \\Deutsches Elektronen-Synchrotron DESY, Notkestra\ss e 85, 22607 Hamburg, Germany}
 \affiliation{The Hamburg Center for Ultrafast Imaging, Luruper Chaussee 149, 22761 Hamburg, Germany}

  \author{Franz X. K\"artner}
   \affiliation{Center for Free-Electron Laser Science CFEL, \\Deutsches Elektronen-Synchrotron DESY, Notkestra\ss e 85, 22607 Hamburg, Germany}
 \affiliation{The Hamburg Center for Ultrafast Imaging, Luruper Chaussee 149, 22761 Hamburg, Germany}
 \affiliation{Physics Department, University of Hamburg, Luruper Chaussee 149, 22761 Hamburg, Germany}

 \author{Angel Rubio}
  \email{angel.rubio@mpsd.mpg.de}
\affiliation{Max Planck Institute for the Structure and Dynamics of Matter,               Luruper Chaussee 149, 22761 Hamburg, Germany}
 \affiliation{European Theoretical Spectroscopy Facility (ETSF)}
 \affiliation{Center for Free-Electron Laser Science CFEL, \\Deutsches Elektronen-Synchrotron DESY, Notkestra\ss e 85, 22607 Hamburg, Germany}
\affiliation{Physics Department, University of Hamburg, Luruper Chaussee 149, 22761 Hamburg, Germany}

\begin{abstract}
The strong ellipticity dependence of high-harmonic generation in gases enables numerous experimental techniques that are nowadays routinely used, for instance, to create isolated attosecond pulses.
Extending such techniques to high-harmonic generation in solids requires a fundamental understanding of the microscopic mechanism of the high-harmonic generation.
Here, using extensive first-principles simulations within a time-dependent density-functional framework, we show how intraband and interband mechanisms are strongly and differently affected by the ellipticity of the driving laser field.
The complex interplay between intraband and interband effects can be used to tune and improve harmonic emission in solids. In particular, we show that the energy cutoff of the high-harmonic plateau can be increased by as much as 30\% using a finite ellipticity of the driving field, opening a new avenue for better understanding and control of HHG in solids based on ellipticity. Also, we demonstrate the possibility to generate, from a single circularly polarized driving field, circularly polarized harmonics with alternating helicity. Our work shows that ellipticity provides an additional knob to experimentally control high-order harmonic generation in solids.
\end{abstract}

\maketitle

Taking advantage of the polarization state of light pulses has recently opened up tremendous, unprecedented opportunities for investigating and controlling strong-field interactions in atomic and molecular gases. The polarization degree of freedom is not only important for studying fundamental physical aspects of light-matter interactions, but a time-varying polarization state \cite{Brixner2004,Kerbstadt2017} underlies numerous spectroscopy and coherent control techniques in attoscience, and it is technologically relevant for tabletop high-harmonic sources in the extreme ultraviolet (XUV) and soft-X-ray spectral regions.

For example, in atomic and molecular gases, attosecond recollision-based physical processes, such as laser-induced electron diffraction  \cite{Blaga2012}, nonsequential double ionization \cite{Niikura2002}, above-threshold ionization \cite{Paulus1998, Kopold2000} and high-harmonic generation (HHG) \cite{Corkum1994,Kopold2000},
are extremely sensitive to small deviations from linear polarization due to the
resulting lateral displacement of the returning electron wavepacket with respect to the parent ion (as nicely accounted for by the standard recollision model of strong-field physics~\cite{PhysRevLett.70.1599,PhysRevLett.71.1994,kuchiev1987atomic}).
The ellipticity-dependence of HHG was recently used to probe the molecular chirality on a sub-femtosecond electronic timescale \cite{Cireasa2015}.
More technologically, this ellipticity sensitivity has been successfully exploited in several gating schemes for the production of \emph{isolated} attosecond XUV pulses, e.g., by polarization gating \cite{Sansone2006} and (generalized) double optical gating \cite{Mashiko2008,Feng2009}.

Coherent steering of the electron wavepacket in a two-dimensional plane using orthogonally polarized two-color laser fields allows to measure the tunnel ionization time and recollision time \cite{Shafir2012}, as well as probing the parent ion with the electron returning under different angles with attosecond precision \cite{Kitzler2005}, which brings intriguing applications in the tomography of atomic or molecular wavefunctions \cite{Shafir2009}. Even more elaborate schemes using counter-rotating circularly polarized laser fields at different wavelengths has lead to the recent demonstration of bright circularly polarized soft-X-ray high-harmonic sources with fascinating spectroscopic applications of magnetic materials using X-ray magnetic circular dichroism (XMCD) (see for reference Refs. \cite{kfir2015generation,Fan2015,Hickstein2015, Chen2016}, and earlier works cited therein).

The use of circularly polarized fields opens the door to producing vortex-shaped photoelectron momentum distributions \cite{Pengel2017} as well as studying spin-polarized electrons created by nonadiabatic tunneling, \cite{Barth2011,Herath2012,Hartung2016,Ayuso2016} attosecond control of spin-resolved recollision dynamics \cite{Ayuso2016}, and investigating ionization dynamics from atoms and molecules via angular streaking ('atto-clock') \cite{EckleNP2008,EckleSci2008,Staudte2009,WuNC2012} using cold target recoil ion momentum spectroscopy (COLTRIMS).

From the above, it is clear that the driver field's ellipticity for strong-field interactions in gases has opened up a plethora of interesting physical phenomena to explore.
In contrast, the role of ellipticity in strong-field interactions in solids remains so far largely unexplored, thus hampering the possibility to exploit or extend some of the above-mentioned experimental techniques to solid-state devices.

The first experimental investigation of the impact of the ellipticity of the driving laser field on HHG from bulk ZnO~\cite{Ghimire2011} showed that the emitted harmonics are less sensitive to ellipticity than harmonics originating from gases. 
However, like in atoms and molecules, circularly polarized light suppresses HHG from this material~\cite{Ghimire2011}.
Solving the semiconductor Bloch equations for a two-band model for ZnO shows that the harmonic yield monotonically decreases with a Gaussian profile with increasing ellipticity ~\cite{PhysRevA.93.043806}. Such an atomic-like monotonic decrease of the harmonic yield with increasing driving laser ellipticity was recently observed experimentally also from rare-earth solids~\cite{ndabashimiye2016solid} and monolayer MoS$_2$~\cite{liu2016high}.
However, a later work on bulk MgO~\cite{you2016anisotropic} reported that, unlike in gases, HHG from bulk crystals can exhibit strongly \emph{anisotropic} ellipticity profiles.
The authors showed that the maximal harmonic yield can, in some cases, be reached not for linear polarization, but for a finite value of the ellipticity $\epsilon$. 
Their experimental results also revealed that, counter-intuitive to previous belief, circularly polarized driver pulses do not always prohibit harmonic generation from bulk crystals.

In order to explain the strongly anisotropic ellipticity dependence of HHG in MgO, You \textit{et al.} proposed a model based on classical \emph{real-space} trajectories in a two-dimensional one-band model including scattering from neighboring atomic sites.
However, their simple picture of pure intraband dynamics is physically incompatible with real-space classical trajectories:
In fact, the adiabatic evolution within one band in momentum space (Bloch oscillations) corresponds to a Wannier-Stark localization in real space,\cite{RossiKuhn} for which electrons localize at different atomic sites of the crystal~\cite{schiffrin2013}, as experimentally observed in semiconductor superlattices~\cite{Lyssenko1997}.
The possibility of maximal harmonic yield at finite ellipticity was proposed for solids, in the regime of semi-metallization of the crystal~\cite{PhysRevB.94.241107}.
As this semi-metallization regime occurs at a much higher intensity than used in the HHG experiments in solids so far, it cannot explain the experimental results of Ref.~\cite{you2016anisotropic}.

Here, we investigate, using an \textit{ab-initio} approach based on time-dependent density-functional theory (TDDFT)~\cite{PhysRevLett.52.997, PhysRevLett.80.1280}, the role of ellipticity in HHG from solids. Simulations are performed for bulk silicon and bulk MgO.
We follow the approach we have recently introduced in Ref.~\onlinecite{OurPRL2017} to describe HHG in solids with full inclusion of electronic band structure and crystal structural effects (see Ref.~\onlinecite{OurPRL2017} and the section Methods for more technical details).

\section{Results}
\subsection{Influence of the driving field's ellipticity}
We start by analyzing the ellipticity dependence of HHG in the case of bulk silicon.
The vector potential acting on the electrons is given by (atomic units are used throughout this paper)
\begin{eqnarray}
 \mathbf{A}(t) = \frac{\sqrt{I_0}c}{\omega}f(t)\Big[\frac{1}{\sqrt{1+\epsilon^2}}\cos(\omega t + \phi)\mathbf{\hat{e}_x}
 +  \frac{\epsilon}{\sqrt{1+\epsilon^2}}\sin(\omega t + \phi)\mathbf{\hat{e}_y}\Big],
\end{eqnarray}
where $I_0$ is the peak intensity inside matter, $f(t)$ the (normalized) envelope, $\omega$ the carrier photon energy, and $\phi$ the carrier-envelope phase (CEP).

We neither account for dephasing nor propagation effects in our simulations, but we found that the recently reported experimental ellipticity profiles of HHG in bulk MgO are well reproduced by our theoretical description (see below), showing the reliability of our theoretical description. Surface effects, as well as light-propagation effects and dissipation via phonons are beyond the scope of the present work.

\begin{figure}[t]
  \begin{center}
    \includegraphics[width=0.7\columnwidth]{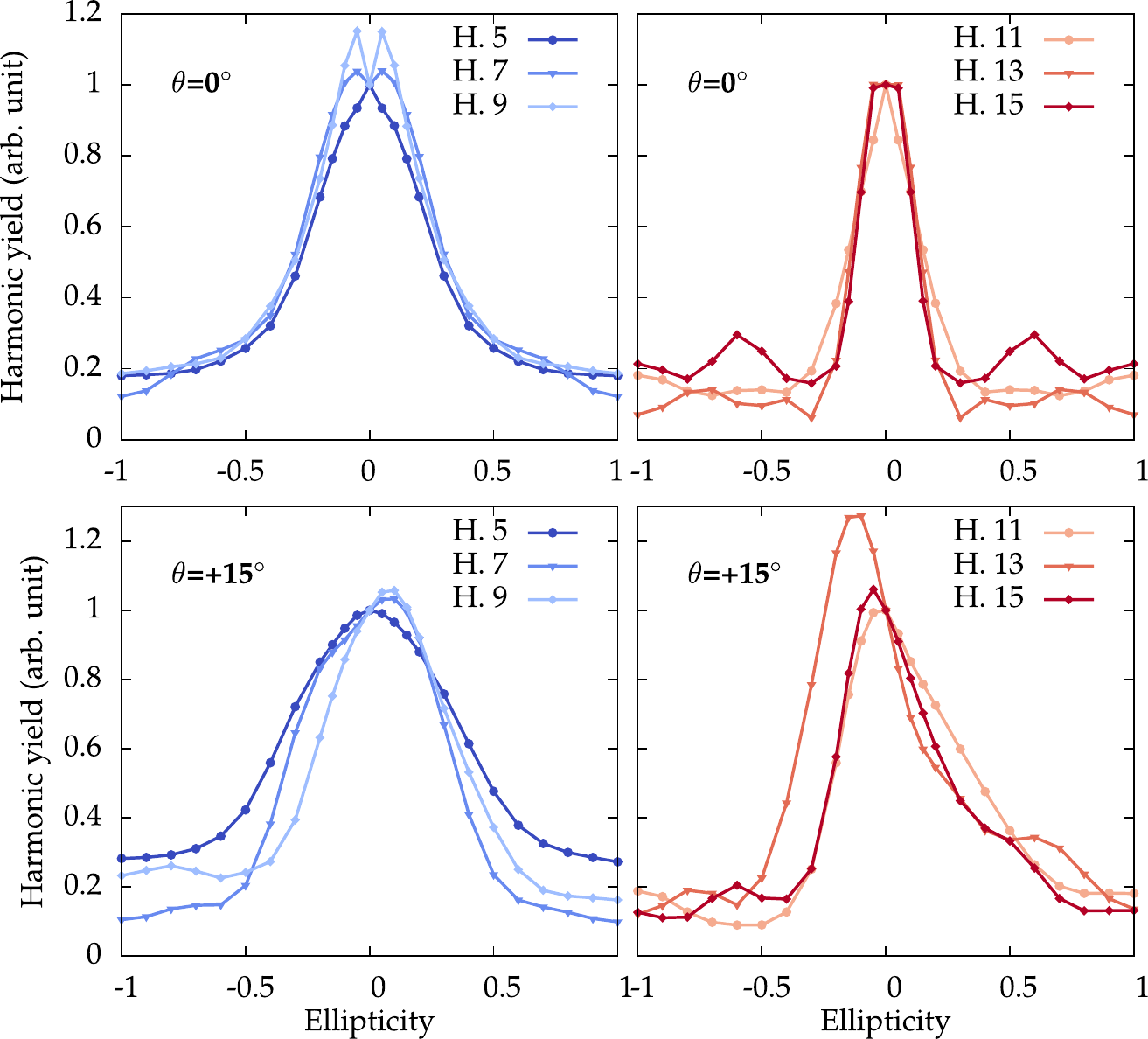}
  \end{center}
  \caption{\label{img:ellipticity} TDDFT simulations of the ellipticity dependence of the various odd harmonics (5 to 15) generated from bulk silicon for laser polarization along the $\overline{\Gamma X}$ direction ($\theta=0^{\circ}$; top panels),
  and for laser polarization rotated by +15$^{\circ}$ around the [001] crystallographic axis (bottom panels). Two distinct responses are observed for harmonics 5 to 9 and for harmonics 11 to 15 (see main text for details)}
\end{figure}

Considering the microscopical mechanism underlying HHG from solids, we note that if the laser field is elliptically polarized with a major axis along a mirror plane of the Brillouin zone (BZ) of the crystal, the left-handed (defined here by negative ellipticity $\epsilon$) and right-handed (positive ellipticity) helicities are equivalent.
This is well understood as the HHG mechanism reflects the symmetries of the BZ~\cite{OurPRL2017}.
Following the same argumentation, if the major axis of the polarization ellipse of the driving field is not aligned with a major axis, we expect an anisotropic profile as left-handed and right-handed helicities will drive electrons into different and non-equivalent regions of the BZ, as experimentally observed recently~\cite{you2016anisotropic}.
Our simulation results, shown in Fig.~\ref{img:ellipticity}, clearly predict an isotropic ellipticity profile for a laser polarization along the $\overline{\Gamma X}$ direction (top panels), whereas an anisotropic profile is found if the major axis of polarization is rotated by $+15^\circ$ around the [001] crystallographic axis (bottom panels).

Interestingly, our results show that for the laser polarization along the $\overline{\Gamma X}$ direction (top panels), the harmonics 5 to 9 exhibit a very similar ellipticity dependence.
In contrast, harmonics 11 to 15 present a different profile but exhibit all a very similar ellipticity dependence.
This puts in evidence that the physical interpretation of Ref.~\cite{you2016anisotropic}, for which classical real-space trajectories can only lead to the same ellipticity dependence for \emph{all} the emitted harmonics, should be revisited, as we will do next here.

Recently, we demonstrated in \cite{OurPRL2017} the possibility to predict spectral regions in the emitted HHG spectra, where the interband contribution is suppressed, from the knowledge of the joint density of states (JDOS).
For the same material and laser parameters used here, it was found that harmonics 5, 7 and 9 do not exhibit a clean odd-harmonic peak structure and appear quite noisy, which is consistent with both interband and intraband mechanisms contributing to HHG.
On the other hand, the harmonics originating mostly from the intraband mechanism (harmonics 11 to 15) were found to have a clean structure~\cite{OurPRL2017}.
From the results displayed in Fig.~\ref{img:ellipticity} (and from the ellipticity of the emitted harmonics, see below), we recover the same grouping of harmonics, based on their ellipticity dependence. This indicates that interband and intraband mechanisms respond differently to ellipticity.

In order to get deeper insight in this interpretation, we have reproduced the same simulations, but for a laser polarization rotated by +15$^{\circ}$ around the [001] crystallographic axis.
Again, our results (see bottom panels in Fig.~\ref{img:ellipticity}) show that harmonics 5 to 9 behave similarly, as they are
all slightly biased toward right-handed helicity, whereas harmonics 11 to 15 are biased toward left-handed helicity. 
This clearly shows that these two groups of harmonics do not have the same physical origin, and  indicates that the two microscopic mechanisms responsible for HHG in solids,
namely the interband and intraband mechanisms, are affected differently by the ellipticity of the driving field.

\begin{figure}[t]
  \begin{center}
    \includegraphics[width=0.65\columnwidth]{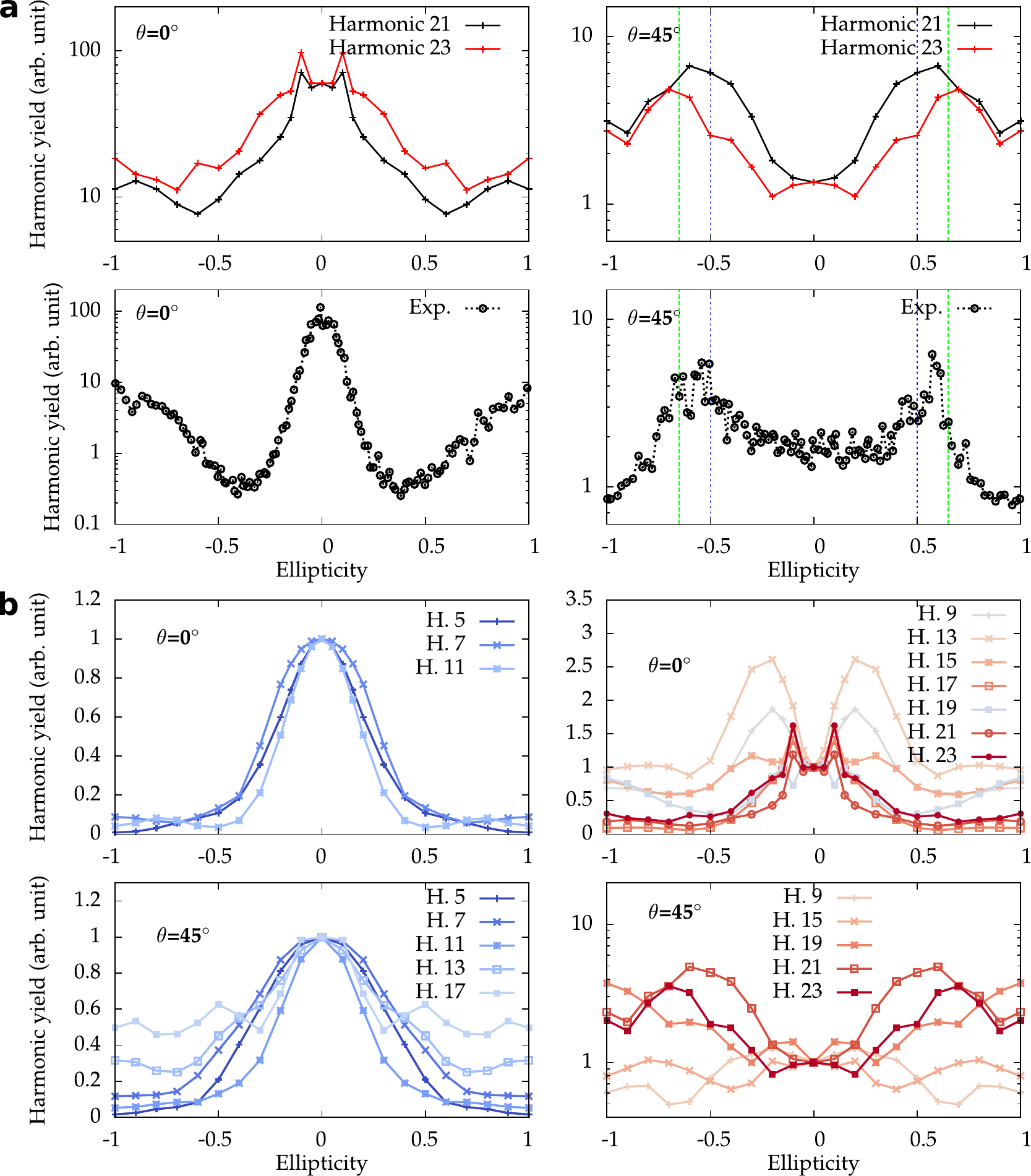}
  \end{center}
  \caption{\label{img:MgO_full} TDDFT simulation of the ellipticity dependence of the harmonic yield for the different harmonics of bulk MgO. (a) Calculated ellipticity dependence of the harmonics 21 (black lines) and 23  (red lines) from bulk MgO (top panels) versus the experimentally observed ellipticity dependence of harmonic 19 taken from Ref.~\onlinecite{you2016anisotropic} (bottom panels). Vertical lines in the right panels showing the case of the Mg-O direction ($\theta=45^\circ$) indicate the positions of $\epsilon=\pm$0.5 and $\epsilon=\pm$0.65. (b) Ellipticity dependence of various harmonics, for the major axis of the polarization ellipse at $\theta=0^\circ$ (top panels) and  $\theta=45^\circ$ (bottom panels). The harmonics are divided into a group of mostly atomic-like (left panels) and non-atomic-like harmonics (right panels), the latter exhibiting a pronounced increase of the harmonic yield at non-zero values of ellipticity.}
\end{figure}

Our conclusions are further supported by simulations for bulk MgO (see Fig.~\ref{img:MgO_full}), which are in qualitative agreement with experiments \cite{you2016anisotropic}.
We found that low-order harmonics exhibit a completely different ellipticity dependence than higher orders, as presented in Fig.~\ref{img:MgO_full}b. 
This reflects well the altered interplay between the interband and intraband dynamics, which are differently affected by the ellipticity.

It is very important to make a close connection to the case of HHG in atoms. Indeed, in some cases~\cite{Ivanov96,Burnett95} an increase of the harmonic yield for an ellipticity $\epsilon$ $\sim$0.1 has been observed, and it was proposed that these harmonics could originate from bound-bound transitions~\cite{Burnett95}. In the case of solids, this scenario would correspond to interband transitions. In the case of silicon (top left panel in Fig.~\ref{img:ellipticity}) and magnesium oxide (top right panel in Fig.~\ref{img:MgO_full}b), we observe such an increase for the harmonics 7 and 9, which have both interband and intraband contributions for our excitation conditions~\cite{OurPRL2017}. Harmonics 11 to 15, which are mainly originating from intraband contributions, do not exhibit such increase. 
This is just an indication of the role of interband transitions that would require further work to see if it is a general effect or specific of this system.

The fact that the two mechanisms depend differently on the ellipticity of the driving electric field can be understood as follows: In the case of harmonic emission from the interband mechanism, the emission only depends on optical transitions between available energy levels. In the simplified ideal case of emission of harmonics by a pure interband mechanism, the electrons only perform transitions, independently of how they are steered by the laser field in momentum space. This means that left-handed and right-handed elliptic polarizations should not contribute differently to the interband mechanism, as in both cases the field strength, and thus the excitation of electrons, is identical. On the other hand, the intraband mechanism directly probes the conduction bands' dispersion (i.e., the group velocity of the electron wavepacket in momentum space). Moreover, any avoided electronic crossing in the band structure can result in diabatic dynamics, whose concomitant harmonic emission depends of how electrons are driven to this avoided crossing. For HHG in solids, the complex interplay between interband and intraband mechanisms leads to a different weight for each harmonic~\cite{OurPRL2017}, and it is therefore natural to find a variety of ellipticity profiles for different harmonic orders of the same crystal, as shown in particular in the bottom panels of Fig.~\ref{img:ellipticity}.

\subsection{Sub-cycle dynamics of excited electrons}

\begin{figure}[t]
  \begin{center}
    \includegraphics[width=0.8\columnwidth]{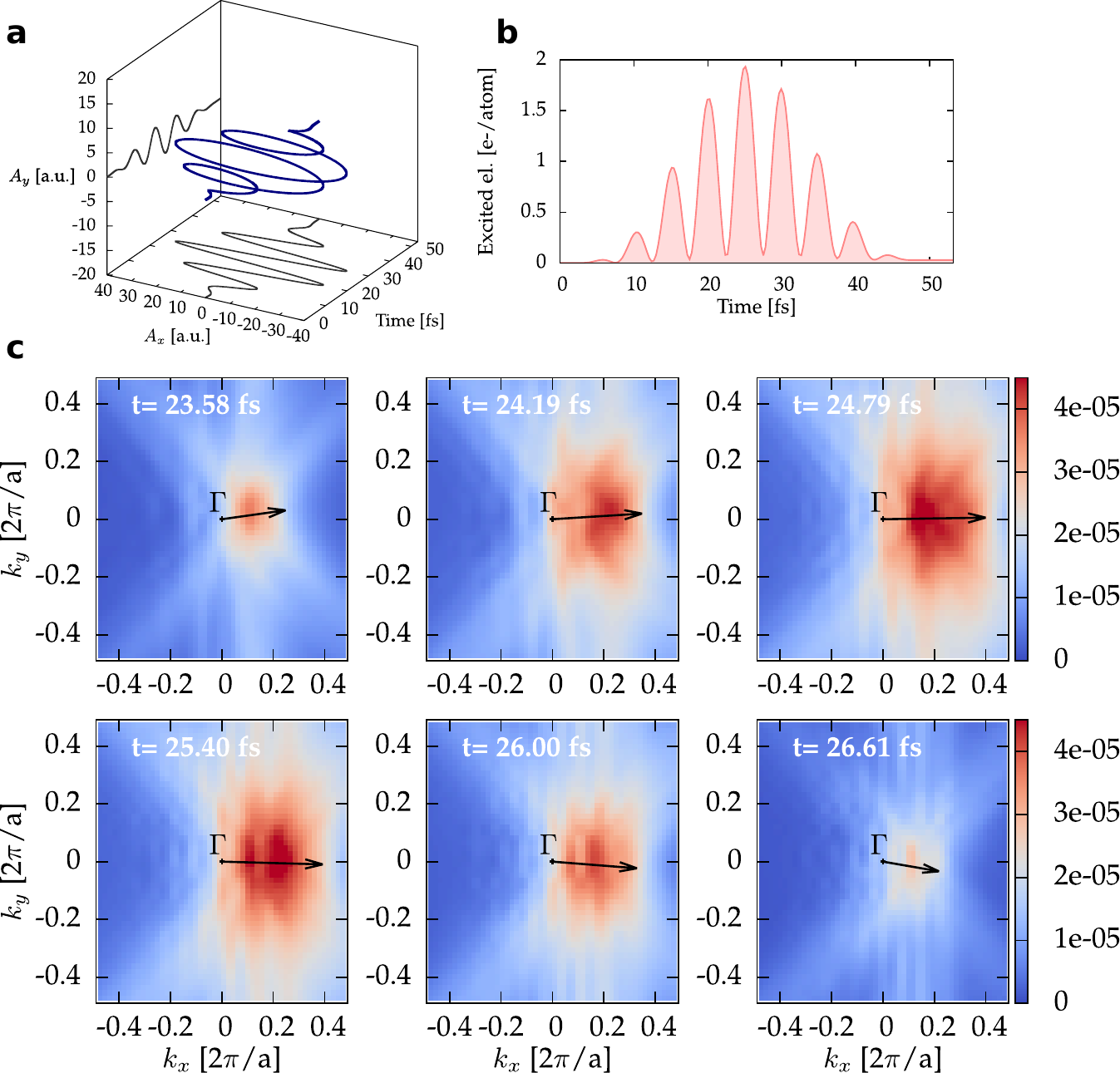}
  \end{center}
  \caption{\label{img:nex_kz_dynamics} TDDFT simulations of the sub-cycle dynamics of the excited electrons around the maximum of the laser vector potential, in the $k_z=0$ plane. a) The left-handed applied vector potential, with an ellipticity of $\epsilon=0.1$. b) Number of electrons excited to the conduction bands during the laser pulse. c) Momentum-space resolved sub-cycle dynamics of the excited electrons. The black arrow indicates the direction and strength of the applied vector potential. The number of excited electrons (displayed as colormap) is computed by projecting the time-evolved wavefunctions on the ground-state Kohn-Sham wavefunctions (see Methods section).}
\end{figure}

In Ref.~\cite{OurPRL2017} we showed that the harmonic yield is enhanced when the interband mechanism is suppressed by band-structure effects. 
We now propose to take advantage of the ellipticity of the laser field to drive the electrons into a specific region of the BZ to enhance HHG.
In order to demonstrate the driving of the electron wavepacket in momentum space by the external laser in a real material, we computed the dynamics of the excited electrons, resolved in momentum space. Our results presented in Fig.~\ref{img:nex_kz_dynamics} show that electrons are excited, starting as soon as the field reaches a critical value (for which a sufficient fraction of valence electrons can tunnel to the conduction bands). The electron wavepacket is then subsequently accelerated by the vector potential of the applied laser, indicated by the black arrow. The region explored by the electrons is dictated by the instantaneous value of the vector potential. Moreover, the various snapshots of the excited electrons show a complex modulation at a sub-cycle time scale, due to the complex band-structure of silicon, which results in many conduction 
bands being involved in the dynamics (A video of the full time evolution of the momentum-resolved sub-cycle dynamics of excited states is provided as Supplementary Video 1).

\subsection{Momentum-space trajectories}

It might of course be tempting to interpret the dynamics of the electron wavepacket in terms of $k$-space trajectories using the so-called ``acceleration theorem''~\cite{RossiKuhn}. 
This has been done, for instance, in~[\onlinecite{PhysRevLett.113.073901,DonkeyKong,Du17}] for few-band models or analytical potentials.
%and it was found to agree with the results of the solution of the one-dimensional time-dependent Schr\"odinger equation with an analytic potential\red{REF}.
However, due to the complexity of the band-structure of even simple semiconductors, such as silicon, involving many bands close to the band gap, the validity of such a simple analysis must be scrutinized.
In particular, the underlying adiabatic approximation implies that interband transitions, level crossings and avoided crossings are neglected.

%Here we compare the TDDFT simulations of the evolution of the excited-electron wavepacket to a description based on momentum-space trajectories, as predicted by the acceleration theorem.
%
The acceleration theorem states that~\cite{RossiKuhn}, under the approximation of an adiabatic evolution, the evolution of the electron wavepacket momentum $\mathbf{k}_e$ is given by
\begin{equation}
 \frac{d\mathbf{k}_e}{dt} = \mathbf{F}(t),
 \label{eq:acceleration_theorem}
\end{equation}
where $\mathbf{F}(t)$ is the force acting on the electron wavepacket. 
Neglecting electron-electron and electron-phonon scattering, this force reduces to the driving electric field. 
More precisely, the above formula (\ref{eq:acceleration_theorem}) is only valid if the electron wavepacket remains in the same band, and does not interact with other electrons or phonons. 
%We note here that the underlying adiabatic approximation implies that interband transitions, level crossings and avoided crossings are neglected.
It is therefore clear that it cannot describe a situation where the interband mechanism dominates over the intraband mechanism.

In a typical HHG experiment, electrons are first excited from the valence to the conduction bands \emph{during the laser pulse}. In particular, the field strength needs to reach a critical value, such that interband Zener tunneling occurs with significant probability~\cite{schiffrin2013}.
Afterwards, depending on its  birth time $t_b$ (adopting the language of the three-step model~\cite{PhysRevLett.70.1599,PhysRevLett.71.1994,kuchiev1987atomic}), an excited electron wavepacket will be driven along different trajectories in momentum space, assuming that no interband transitions take place. 
Assuming the validity of the acceleration theorem (i.e., an adiabatic evolution), the trajectory of the center-of-mass of an electron wavepacket is thus given by
\begin{equation}
 \mathbf{k}_e(t) = \mathbf{k}_e(t_b) - \frac{1}{c}\left(\mathbf{A}(t)-\mathbf{A}(t_b)\right)\,,
 \label{eq:trajectories}
\end{equation}
where $t_b$ is the birth time of the electron wavepacket, i.e., the moment at which it is created.
This time accounts for the fact that in HHG experiments reported so far for bulk crystals, the electrons are excited by interband transitions during the laser pulse.
In many previous works, however, the electron wavepacket was usually assumed to already exist before the pulse arrives (for $t_b\rightarrow -\infty$), then Eq.~(\ref{eq:trajectories}) reads
\begin{equation}
 \mathbf{k}_e(t) = \mathbf{k}_e(t_b) - \frac{1}{c}\mathbf{A}(t).
 \label{eq:trajectories_textbook}
\end{equation}
Assuming now, for the sake of argument and illustration, that an electron wavepacket can be created at any birth time, we obtain for each instant in time $t$ a set of positions ($k_x,k_y$), corresponding to all wavepacket created at all previous birth times $t_b$.
As the minimal (direct) band gap of silicon is located at the $\Gamma$ point, we have $\mathbf{k}_e(t_b)=0$.

\begin{figure}[t]
  \begin{center}
    \includegraphics[width=0.8\columnwidth]{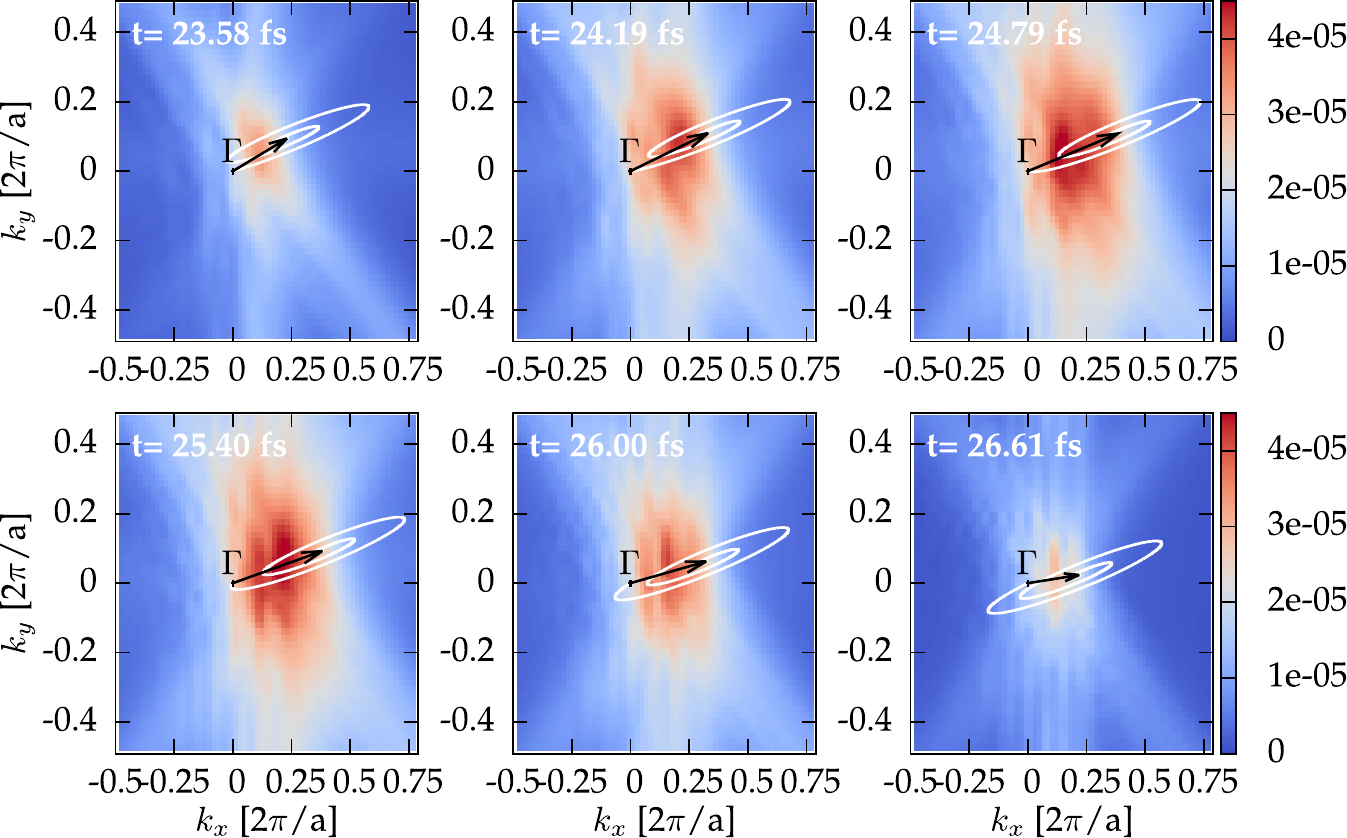}
  \end{center}
  \caption{\label{img:nex_kz_dynamics_traj} Comparison of the time-evolution of the excited electron wavepacket in momentum space $N_{\mathrm{ex}}(\mathbf{k};t)$ computed from TDDFT with possible positions predicted by the acceleration theorem. The black arrow indicates the direction and strength of the vector potential. For each time $t$, the white curve corresponds to all possible center-of-mass positions of the wavepacket, for all possible birth times $t_b$ of the wavepacket. The major axis of the polarization ellipse is rotated by $\theta=+15^\circ$ around the [001] crystallographic axis.}
\end{figure}

Inspecting the time-evolution of the number of excited electrons shown in Fig.~\ref{img:nex_kz_dynamics}b, it is clear that under our excitation conditions, most of the excited electrons are virtually excited electrons, whose number returns almost to zero after each half cycle.
We also note that a more elaborate model should also take into account a critical value for the electric field $E_c$. 
Below this value, no significant portion of the valence electrons are really excited into the conduction bands. By choosing such a value, one restricts the values of the birth time to the cases for which $|E(t_b)|>E_c$.
Here we do not attempt to propose such an elaborate model, in particular because the critical electric field is not a well-defined quantity, and moreover, such a sophisticated model would only remove some of the possible birth times, hence not changing drastically the conclusions drawn from our simple trajectory analysis. 
Assuming that all prior times are possible times of birth for the wavepacket, we obtain the trajectories, as shown in Fig.~\ref{img:nex_kz_dynamics_traj} for the case of the major axis of the polarization ellipse being rotated by $\theta=+15^\circ$ around the [001] crystallographic axis.
In this case, as well as in all cases we investigated, we found that the trajectories obtained from the acceleration theorem agree poorly with our \textit{ab-initio} TDDFT results.
Indeed, the acceleration theorem predicts possible positions of the wavepacket in a wider region of the BZ than actually explored by the electrons, according to our \textit{ab-inito} simulations. 
This shows that neglecting the interband dynamics is not valid for bulk silicon and for our excitation conditions.
Overall, this indicates the breakdown of the simple models used in the literature for explaining HHG from solids in a pure trajectory picture.

\subsection{Ellipticity-based HHG cutoff extension}

The energy cutoff of HHG spectra has always been of main importance for technological applications. In solids, this cutoff depends on the maximum peak of the driving electric field,~\cite{Ghimire2011} as well as on the polarization direction of the driving field, even in cubic materials.~\cite{OurPRL2017} We now show that the cutoff energy also depends on the ellipticity of the driving field and that, in contrast to gases, it can even be increased for finite ellipticity in some cases.
In order to show the effect of ellipticity on the HHG cutoff, we exploit the case of HHG from bulk MgO with a laser polarization along the Mg-O bond. It was found experimentally in \cite{you2016anisotropic}, that an ellipticity of $\epsilon=0.65$ results in an increase of the harmonic yield of bulk MgO by almost one order of magnitude for one of the highest harmonics (19th order). This is well reproduced by our TDDFT simulations (see Fig.~\ref{img:MgO_full}a).

\begin{figure}[t]
  \begin{center}
    \includegraphics[width=0.7\columnwidth]{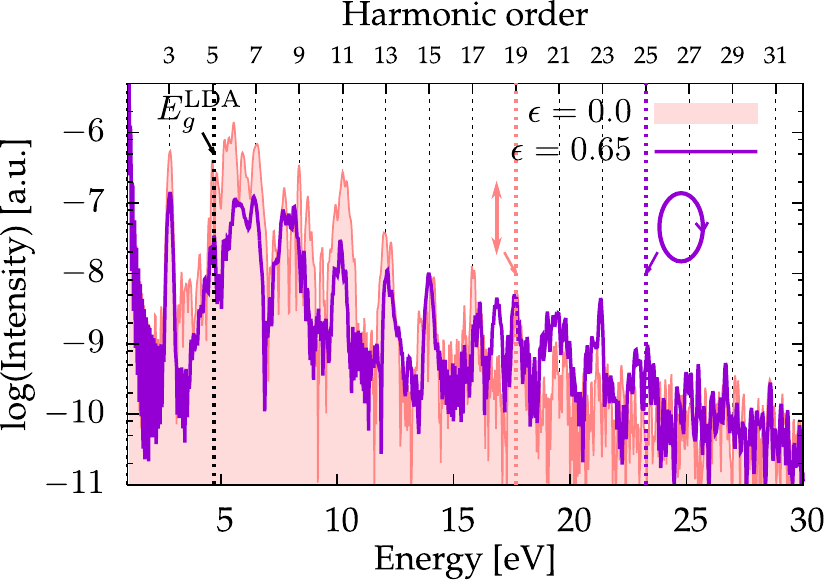}
  \end{center}
  \caption{\label{img:Comp_polarisation} Calculated HHG spectra from bulk MgO for a laser linearly polarized and a laser elliptically polarized ($\epsilon=0.65$). 
  In both cases, the major axis of the polarization ellipse along the $\overline{\Gamma K}$.
  The dashed lines indicate the position of the cutoff energy ($E_c$).
  }
\end{figure}

Our results, reported in Fig.~\ref{img:Comp_polarisation}, show that when the ellipticity of the laser is changed from linear polarization ($\epsilon=0$) to the ellipticity that maximizes both experimental and theoretical harmonic yields of the highest harmonics ($\epsilon=0.65$; see Fig.~\ref{img:MgO_full}), the cutoff energy for the HHG is increased by up to 30\%. 
Thus, our results clearly show that it is possible to strongly modify, and even increase the cutoff energy, by changing the ellipticity of the incoming laser from linear polarization to some finite ellipticity.
This increase of the cutoff is even more impressive, considering that the maximum field at finite ellipticity is a factor $\frac{1}{\sqrt{1+\epsilon^2}}$ (=0.84 for $\epsilon=0.65$) smaller than the field strength for linear polarization.
Therefore, assuming a linear scaling of the cutoff in field strength, we should have found an energy cutoff around 15\,eV, i.e., the 15th harmonic. We instead obtain harmonics up to the 25th harmonic from our first principles simulations.

Our findings highlight that the HHG energy cutoff is not only dictated by the incoming laser field strength but is, in fact, strongly affected by the potential energy landscape felt by the electrons, i.e., the part of the band structure explored by the electrons driven by the strong laser field.
From this perspective, it appears that the laser polarization direction as well as the ellipticity are natural tools to coherently steer electrons inside the BZ, thereby controlling and optimizing the HHG cutoff energy from bulk crystals.

\subsection{Ellipticity and helicity of the emitted harmonics}

In Ref.~[\onlinecite{langer2017symmetry}], it was speculated that it could be possible to generate circularly polarized high-order harmonics from a solid driven by a \emph{single-color} circularly polarized driving field.
It is clear that this could lead to new and simpler spectroscopy techniques such as XMCD for studying magnetic materials \cite{Fan2015,Hickstein2015,Chen2016} compared to the bi-color counter-rotating driver fields used in the gas case.

\begin{figure}[t]
  \begin{center}
    \includegraphics[width=\columnwidth]{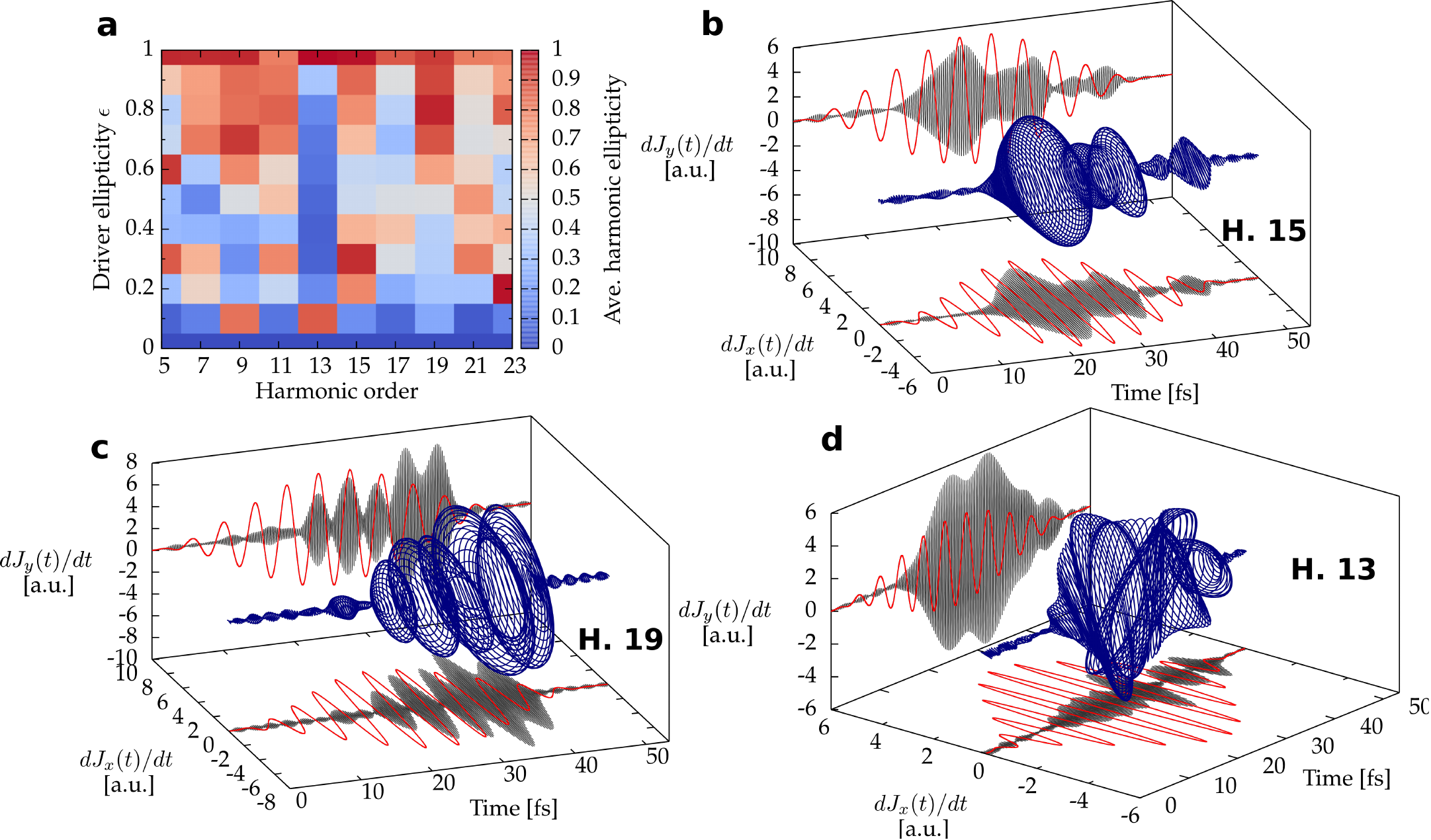}
  \end{center}
  \caption{\label{img:ellipticity_harmonics} Ellipticity of the emitted harmonics for bulk MgO. a) Calculated average ellipticity of the harmonics emitted for the major axis along the $\overline{\Gamma X}$ direction, versus the ellipticity of the driving field. b) Evolution of the time-derivative of the electronic current, bandpass filtered around the 15th harmonic (see Methods section), for circular polarization ($\epsilon=1$). c) Same as b, but for the 19th harmonic. d) Same as b) and c), but for the 13th harmonic and $\epsilon=0.6$.}
\end{figure}

We evaluate now how the average ellipticity of the emitted harmonics (see Methods section) depends on the ellipticity of the driving field.
Our results, presented for bulk MgO  in Fig.~\ref{img:ellipticity_harmonics}a, show that even if the ellipticity (averaged over the pulse duration) of the emitted harmonics does not exactly reproduce the ellipticity of the driving field, there is a clear general trend that the ellipticity of the harmonics increases with increasing ellipticity of the driving field.
Only the 13th harmonic exhibits an average ellipticity close to zero for all driver ellipticities in Fig.~\ref{img:ellipticity_harmonics}a. However, as Fig.~\ref{img:ellipticity_harmonics}d reveals, the ellipticity of the 13th harmonic is simply averaging out to zero due to the time-varying rotation of the polarization ellipse.

In order to get more insight, and to check if it is possible to generate circular harmonics with a single circular laser pulse, we computed the evolution of the time-derivative of the electronic current, filtered in frequency around certain harmonics.
For a circularly polarized driving field, our results (see Fig.~\ref{img:ellipticity_harmonics}b and c) show clearly that the emitted harmonic fields are also mainly circularly polarized.
This result demonstrates the possibility to generate circularly polarized high-order harmonics from a single-color circularly polarized driver pulse in solids,
opening up the door to future investigations of magnetic materials. 
This is a major difference to the HHG in gases, where single circularly polarized driver pulses cannot generate harmonics.

A deeper analysis of the results (a video is provided as Supplementary Video 2) also revealed that the harmonics obtained by the circular driver have alternating helicities, similarly to what has been reported previously reported in the case of atoms for bi-color counter-rotating driver fields~\cite{kfir2015generation,Fan2015}.

\begin{figure}[t]
  \begin{center}
    \includegraphics[width=\columnwidth]{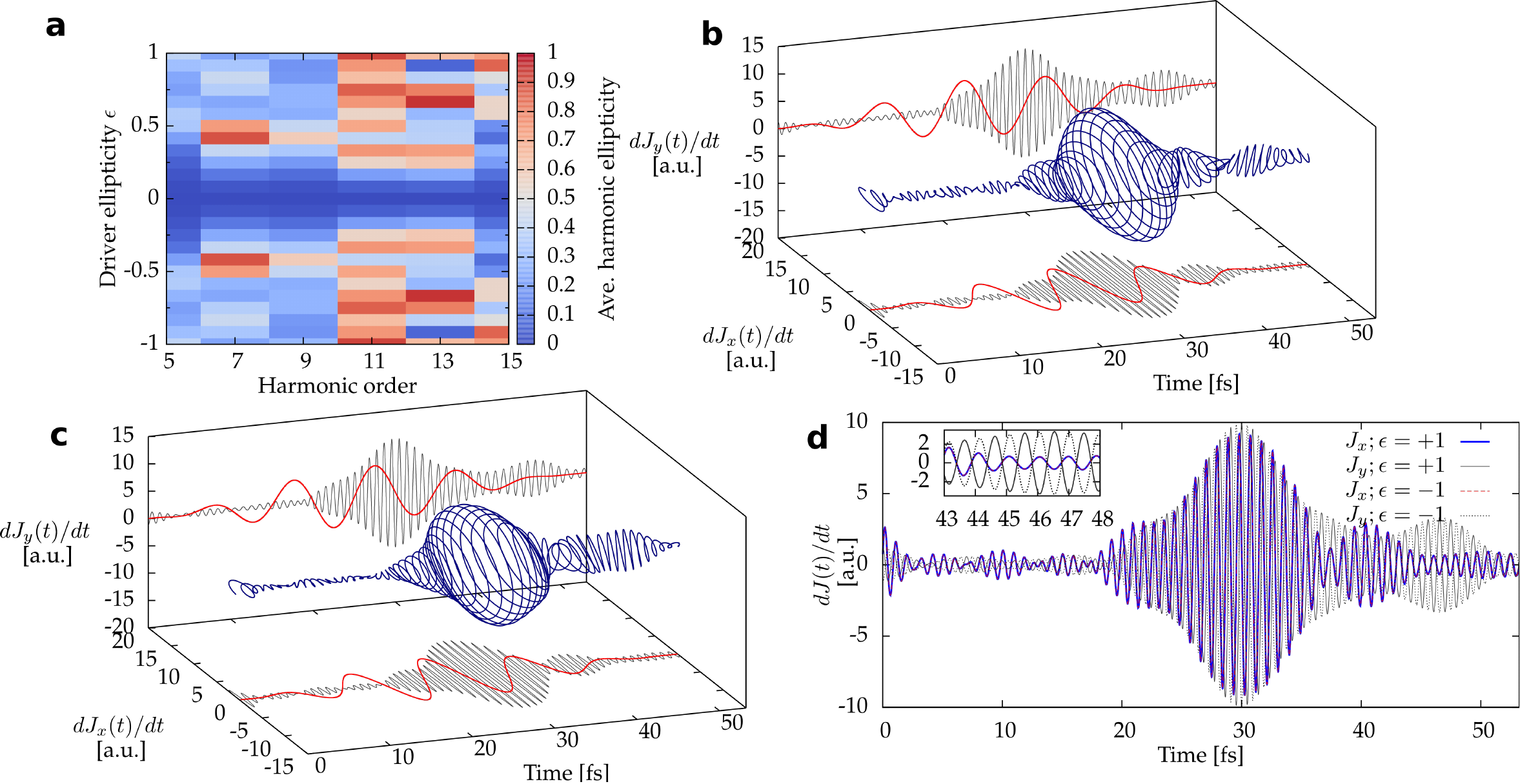}
  \end{center}
  \caption{\label{img:ellipticity_harmonics_Si} Ellipticity of the emitted harmonics for bulk silicon. a) Same as Fig.~\ref{img:ellipticity_harmonics}a but for bulk silicon. b) and c) Evolution of the time-derivative of the electronic current, bandpass filtered around the 11th harmonic for left-handed (b) and right-handed (c) circular polarization ($\epsilon=\pm1$). d) Comparison of the left-handed and right-handed driven 11th harmonic. The $x$-component is found to be identical for the two cases, whereas the $y$ components have opposite phase, showing that they have flipped helicity. }
\end{figure}

We also evaluated the average ellipticity of the emitted harmonics in the case of silicon (see Fig.~\ref{img:ellipticity_harmonics_Si}a).
Interestingly, we recover the same classification of harmonics obtained from the ellipticity profiles, but here according to the ellipticity of the emitted harmonics.
Our results therefore indicate that the generation of harmonics by the interband and intraband mechanisms might lead to distinct ellipticity for the emitted harmonics.
Overall, this is yet another proof that interband and intraband mechanisms are affected differently by the ellipticity of the driving field.

Finally, we investigated the possibility of controlling the helicity of the emitted high-order harmonics. 
We considered in particular the case of the 11th harmonic from bulk Si, shown in Figs.~\ref{img:ellipticity_harmonics_Si}b and ~\ref{img:ellipticity_harmonics_Si}c, respectively, for left- and right-handed circularly polarized driver pulses. 
Our results, shown in Fig.~\ref{img:ellipticity_harmonics_Si}d, clearly demonstrate the possibility of controlling and flipping the helicity of the emitted harmonics by changing the helicity of the driver field.

%%%%%%%%%%%%%%%%%%%%%%%%%%%%%%%%%%%%%%%%%%%%%%%%%%%%%%%%%%%%%%%%%%%%%%%%%%%%%%%%%%%%%%%%%%%%%%%%%%%%%%%%%%%%%%%%%%%%%%%%%%%%%%%%%%
\section{Discussion}
In summary, we have investigated the role of ellipticity of the driving laser field on HHG from solids.
We have shown that the harmonics of different order are not affected equally by the ellipticity and that the symmetries of the Brillouin zone are reflected in the ellipticity profiles.
This can be explained by the fact that the interband and intraband mechanisms exhibit a different ellipticity dependence.
Moreover, we found that the energy cutoff of the HHG spectra can be strongly modified and even increased when changing the ellipticity of the driving field.
Based on our results, we propose to custom-tailor and enhance the HHG in solids by driving the electrons inside the potential energy landscape into specific regions of the Brillouin zone, in particular using laser fields with a temporally evolving polarization state using modern spatial light modulator technology.
Finally, we have demonstrated the possibility of generating (nearly) circularly polarized high-order harmonics with alternating helicities from a single-color circularly polarized laser field, and to control the helicity of these harmonics.
Our results open new avenues for new ellipticity-based experimental techniques based on high-order harmonic generation, in which solids will play the predominant role.

\section{Methods}

\subsection{TDDFT simulations}

The evolution of the wavefunctions and the evaluation of the time-dependent current is computed by propagating the Kohn-Sham equations within TDDFT, as provided by the Octopus package~\cite{C5CP00351B}, in the adiabatic local-density approximation (LDA).\cite{RevModPhys.74.601}  We employ norm-conserving pseudo-potentials.
The HHG spectrum is directly obtained from the total electronic current $\mathbf{j}(\mathbf{r},t)$ as
\begin{equation}
 \mathrm{HHG}(\omega) = \left|\mathrm{FT}\Big(\frac{\partial}{\partial t}\int d^3\mathbf{r}\, \mathbf{j}(\mathbf{r},t)\Big)\right|^2,
\end{equation}
where $\mathrm{FT}$ denotes the Fourier transform.

\subsection{Simulations of HHG from bulk silicon}

All calculations for bulk silicon were performed using the primitive cell of bulk silicon, using a real-space spacing of 0.484 atomic units. We consider a laser pulse of 25-fs duration at full-width half-maximum (FWHM) with a sin-square envelope, and the carrier wavelength $\lambda$ is 3000\,nm, corresponding to $\omega=0.43$\,eV.
Except for the calculation of the electron dynamics (see below), we employed an optimized 28$\times$28$\times$28 grid shifted four times to sample the BZ, and we used $I_0=10^{11}$W\,cm$^{-2}$ (corresponding to a peak intensity in matter of $3.4\times 10^{12}$\,W\,cm$^{-2}$ for an optical index of $\sim 3.4$).  
We use the experimental lattice constant $a$ leading to a LDA band gap of silicon of 2.58\,eV. In all our calculations, we used a carrier-envelope phase (CEP) of $\phi=0$. We checked (see Supplemental Material) that the CEP has almost no effect on the ellipticity dependence of HHG in solids for the pulse duration considered here.

\subsection{Simulations of HHG from bulk magnesium oxide}

We also performed calculations for bulk MgO, which has a zinc-blende crystal structure. 
We used a real-space spacing of $0.3$ atomic units and an optimized 28$\times$28$\times$28 grid shifted four times to sample the BZ.
We use a carrier wavelength $\lambda$ of 1333\,nm, corresponding to a carrier photon energy of 0.93\,eV to match the experimental conditions used in Ref.~\cite{you2016anisotropic}.
The experimental peak intensity in vacuum is $\sim$ $10^{13}$ W\,cm$^{-2}$. The corresponding transmitted peak intensity in matter is therefore  $\sim$ $9\times10^{12}$ W\,cm$^{-2}$, taking the experimental refractive index of bulk MgO as $1.7175$,~\cite{stephens1952index} for the considered wavelength.
We note that within the local-density approximation (LDA), the band gap of MgO is found to be $E^{\mathrm{LDA}}_g=4.72$\,eV, which strongly  underestimates the experimentally observed band gap of  $E^{\mathrm{exp}}_g=7.83$\,eV~\cite{Whited19731903}.
Therefore, we use $I_0=3\times10^{12}$\,W\,cm$^{-2}$ (corresponding to a peak intensity in matter of $5.3\times 10^{12}$\,W\,cm$^{-2}$ for $n=1.7175$), in order to generate a similar number of harmonics as measured experimentally, to allow a comparison.
In Ref.~\onlinecite{you2016anisotropic}, authors used a 50-fs FWHM laser pulse. In our simulations, we used instead a shorter laser pulse of 25-fs FWHM, in order to make the calculations numerically tractable.  
We found (see Supplemental Material) that the HHG spectra from bulk MgO are very similar for both 25-fs and 50-fs pulse durations. 

\subsection{Definition of the harmonic yield}
\label{sec:yield}

For each odd harmonic, we define the harmonic yield by integrating the HHG spectrum over the energy region defined by the two neighboring even harmonics, such as the harmonic yield of the $n$-th (odd) harmonic is given by
\begin{equation}
 I_{HH,i}(n) = \int_{(n-1)\omega}^{(n+1)\omega} \mathrm{HHG}_i(\omega') d\omega',
\end{equation}
where $\omega$ is the frequency of the laser. If specified, the subscript $i$ indicates that the yield is computed by only taking into account the $i$-component ($i$=x,y) of the total electronic current.

\subsection{Sub-cycle dynamics of the excited electron in momentum space}

The simulations of the sub-cycle dynamics of the excited electrons in momentum space were performed for an intensity of the laser of $I_0=5\times10^{11}$ W\,cm$^{-2}$. The ellipticity is taken as $\epsilon=0.1$.
%, and the major axis of the polarization ellipse is aligned along the $\overline{\Gamma X}$ high-symmetry direction. 
The $t=0$ time corresponds to the switch-on of the laser pulse. In these simulations, we employed a $27\times27\times27$ $\mathbf{k}$-point grid, shifted four times, to get the $k_z=0$ plane in our $\mathbf{k}$-point grid.
The total number of excited electron is defined by projecting the time-evolved wavefunctions ($|\psi_n(t)\rangle$) on the basis of the ground-state wavefunctions ($|\psi_{n'}^{\mathrm{GS}}(t)\rangle$) 
\begin{equation}
 N_{\mathrm{ex}}(t) = N_e - \frac{1}{N_\mathbf{k}}\sum_{n,n'}^{\mathrm{occ.}}\sum_{\mathbf{k}}^{\mathrm{BZ}} |\langle \psi_{n,\mathbf{k}}(t) | \psi_{n',\mathbf{k}}^{\mathrm{GS}} \rangle|^2,
\end{equation}
where $N_e$ is the total number of electrons in the system, and $N_{\mathbf{k}}$ is the total number of  $\mathbf{k}$-points used to sample the BZ. The sum over the band indices $n$ and $n'$ run over all occupied states.
The momentum-resolved excited electron distribution, as shown in Figs.~\ref{img:nex_kz_dynamics} and ~\ref{img:nex_kz_dynamics_traj} is defined here as
\begin{equation}
 N_{\mathrm{ex}}(\mathbf{k};t) =  \frac{1}{N_\mathbf{k}}\Big(N_e - \sum_{n,n'}^{\mathrm{occ.}}|\langle \psi_{n,\mathbf{k}}(t) | \psi_{n',\mathbf{k}}^{\mathrm{GS}} \rangle|^2\Big).
\end{equation}

\subsection{Average ellipticity of the emitted harmonics}
For the case of the driving field being polarized in the $x$-$y$ plane, with the major axis of the polarization ellipse along the $x$-axis, we define the average (over the pulse duration) ellipticity of $n$-th harmonic as
\begin{equation}
 |\epsilon(n\omega)| = \left|\frac{\tilde{E}_y(n\omega)}{\tilde{E}_x(n\omega)}\right| = \sqrt{\frac{I_{HH,y}(n\omega)}{I_{HH,x}(n\omega)}},
\end{equation}
where $\omega$ is the frequency of the driving field, $\tilde{E}_i(n\omega)$ is the strength of the $n$-th harmonic electric field along the direction $i=x,y$, and $I_{HH,i}(n\omega)$ the harmonic yield (as defined in Methods Section~\ref{sec:yield}), directly obtained from the HHG spectra.
However, in some cases, one has to assume that the major axis of the polarization ellipse for the emitted harmonics is along the $y$-axis to get an ellipticity between 0 and 1.
Therefore, we use 
\begin{equation}
 |\epsilon(n\omega)| = \min\left( \sqrt{\frac{I_{HH,y}(n\omega)}{I_{HH,x}(n\omega)}};  \sqrt{\frac{I_{HH,x}(n\omega)}{I_{HH,y}(n\omega)}}\right)
 \label{eq:ellipticity_harmonics}
\end{equation}
to evaluate the ellipticity of the emitted harmonics.
We note that this can only provide an estimate of the ellipticity of the emitted harmonics, as we use here the harmonic yield obtained by integrating the HHG spectra between the two neighboring even harmonics.

\section{Data Availability}
The data that support the findings of this study are available from the corresponding authors upon request, and will be deposited on the NoMaD repository.

\section{Code Availability}
 The OCTOPUS code is available from http://www.octopus-code.org.

%%%%%%%%%%%%%%%%%%%%%%%%%%%%%%%%%%%%%%%%%%%%%%%%%%%%%%%%%%%%%%%%%%%%%%%%%%%%%%%%%%%%%%%%%%%%%%%%%%%%%%%%%%%%%%%%%%%%%%%%%%%%%%%%%%
\section{Acknowledgments}
%%%%%%%%%%%%%%%%%%%%%%%%%%%%%%%%%%%%%%%%%%%%%%%%%%%%%%%%%%%%%%%%%%%%%%%%%%%%%%%%%%%%%%%%%%%%%%%%%%%%%%%%%%%%%%%%%%%%%%%%%%%%%%%%%%
We acknowledge financial support from the European Research Council (ERC-2015-AdG-694097), COST Action MP1306 (EUSpec).
N. T.-D. thanks M. J. T. Oliveira for providing some of the pseudopotential files.
 F.X.K. and O.D.M. thank N. Klemke and Y. Yang for helpful discussions and acknowledge support by the excellence cluster 'The Hamburg Centre of Ultrafast Imaging-Structure, Dynamics and Control of Matter at the Atomic Scale' and the priority program QUTIF (SPP1840 SOLSTICE) of the Deutsche Forschungsgemeinschaft.

\section{Author contributions}
N.T.-D. and A.R. conceived and designed the project. N.T.-D. carried out the code implementation and the numerical calculations. O.D.M. and N.T.-D. worked out the models and the comparison with experiments. All authors participated in the discussion of the results and contributed to the manuscript.

\bibliography{bibliography}

\end{document}